# Long-Range Time-Synchronisation Methods in LoRaWAN-based IoT

Timothy Jones, Khondokar Fida Hasan
Queensland University of Technology

## Abstract

The IoT (Internet of Things) is a network comprised of internet-connected devices and sensors that remotely report and sense data autonomously. It is forecasted that this network of devices will span 20.4 billion by 2020 (Gartner, 2016). These devices operate through data transmissions, and as such, require constant and accurate time synchronisation. A variety of network time protocols exist to serve this purpose, and all aim to accurately synchronise the time through various methods and schemes to adjust the internal clock consistently. This is of particular importance within the field of IoT, as the sole purpose of an IoT node is to sense and report data periodically. If time is not accurately synchronised on a regular basis, a lack of coordination arises within nodes situated within a distributed system, and the reported data will not provide a complete picture of the situation under monitoring.

LoRa (Long-Range) is an LPWAN (low-power wide-area network) protocol that is part of the IoT family that focusses on long-range communication of up to 14km, albeit with delay-inherent transmissions. Three IoT-based time synchronisation methodologies are analysed, and their efficacy measured through a systematic critical literature review. These include a GNSS-based method, an off-the-shelf GPS hardware resampling method, and the LongShoT method, within the context of vehicular ad-hoc networks (VANET), wireless sensor networks (WSN), and long-range wide area network (LoRaWAN) respectively. Although two of the three methods are not LoRaWAN-specific, the findings obtained from the research are applied to the context of LoRa in the proposed methodology.

A methodology for selecting a time synchronisation methodology with regards to LoRa specifically is posited, whereby each requirement of synchronisation objective, energy consumption and costs, scenario and security analysis, application requirements, microcontroller requirements and transceiver requirements are taken into consideration. These are then followed by a fine-grain approach to the selection of a particular time-sync method. The resultant methodology may not only have implications in the field of research, where practitioners may adopt this literature review as a baseline understanding of time synchronisation methods and obstacles encountered toward LoRa, however developers of applications for the LoRaWAN may adapt the analysed methods outlined within.

**Table of Contents**

## List of Figures

1.Introduction

1.1 Diversity of the IoT Network

The IoT network is defined as the integration of connectable devices and sensors to enable remote monitoring, creating a network of networks of autonomous objects (Tirado-Andrés et. al, 2019). By 2020, it is forecasted that the amount of worldwide connected IoT devices will approximate 20.4 billion (Gartner, 2019). Statista predicts that the number of IoT connected devices is expected to nearly quadruple to 75.44 billion by the year 2025 (Statista, 2016). More and more IoT devices are being created that are cost effective and provide convenient functionality for even consumers to use within their own home. These devices aim to send small amounts of data periodically, whilst tolerating delays in traffic.

There currently exists six main IoT protocols used in data transmission from IoT nodes to their gateway(s): Zigbee, Bluetooth Low-Energy (BLE), WirelessHART, LoWPAN, LoRaWAN and Z-Wave. Out of the previously mentioned figure of 20.4 billion IoT devices connected worldwide, Bluetooth is the most widely used IoT protocol by far at an expected 8.45 billion devices in use by 2020 (Statista, 2017). This figure comprises approximately 41% of the forecasted amount of all IoT devices and can be found in everyday devices such as laptops and mobile phones, to in-car units and headphones. This is followed by Zigbee, which is most notably used within short-range sensor networks and home devices such as Amazon Echo, Nest Thermostats, and Philips Hue light bulbs (Stables, 2019). Most of these protocols - with the exception of Bluetooth - are part of the IEEE 802.15.4 standard, and are situated within the Data and Physical Layer of the five-layer OSI model to serve data to the application layer in various formats (Figure 1).

**Figure 1.** The IoT Stack (Heđi et. al 2017)

These common standards short-range are not significantly far away from their respective gateway unlike LoRaWAN, therefore time synchronisation is not a significantly impacted, especially with the unrestricted-size data transmissions being constantly relayed from node to gateway.

1.2 Time Synchronisation and its Importance in Networking

Time synchronisation (TS) aims to accurately synchronise time through various network protocols that adjust the internal clock to attain consistency. It is important in any network due to accurate timestamping being required, for example in geophysical system monitoring, or wireless communication protocols (Gonzalez, 2015). Gonzalez (2015) outlines that although a TS protocol synchronises time, the internal clock itself is not modified. This is because the internal clock is a count in the hardware of clock source pulses. The protocol only computes the local clock shift and adds the calculated shift to the internal clock's value to determine the actual time. TS in the context of IoT is defined as the way an IoT device adjusts its internal clock in order to align with the clocks of other devices in the network (Tirado-Andrés et. al, 2019).

### 1.3 The LoRaWAN Specification

LoRa was originally created as a solution for long-distance communication that does not require a high sampling rate. This allows the protocol to achieve scalability, cellular architecture, and a central coordination function. Each LoRa radio is based on long-range radios created by the Semtech Corporation (Rizzi et. al, 2017). The specification is built around the nature of long-range transmissions with a range of up to 14km. This is achieved through low duty cycles through higher network transmission delays in comparison to other IoT technologies.

A LoRaWAN node typically consists of radio hardware different to that of the gateway. The end device specifically does not contain a hardware counter and instead requires an interrupt line signalling the host microcontroller when a transaction completes (Ramirez et. al, 2019). The gateway itself has the capacity to decode up to 8 LoRa packets simultaneously. A 32-bit hardware counter contained within a typical SX1301 chip has a resolution of 1μs, which can also be attached to an external GPS 1PPS signal. This allows for either a scheduled sending time or a hardware timestamp of incoming packets from the radio.

The LoRa protocol is based in the Physical (PHY) and Datalink layers. It uses chirp spread spectrum (CSS) in order to operate in unlicensed bands. At the receiving end, the signal is downmixed to revert it to the baseband, which transforms it into unchirped through multiplication via a reference chirp generated locally (Ramirez et. al, 2019). Typical bandwidths range from 125-250kHz. The LoRaWAN message format including the PHY data is displayed below in Figure 2.

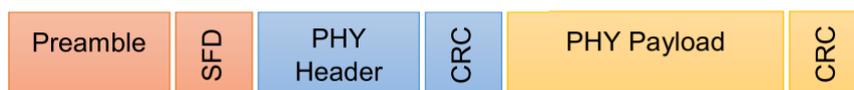

**Figure 2.** LoRaWAN message format (Ramirez et. al, 2019)

### 1.4 The Importance of Time Synchronisation In LoRa

Without Time Synchronization, two main issues arise – an absence of consistency and coordination of devices throughout the distributed system and a lack of a complete picture of the scenario under monitoring from the information from different nodes (Ramirez et. al, 2019; Tirado-Andrés et. al, 2019). This is especially apparent in LoRaWAN, where each node is a part of a star-of-stars topology, where end devices communicate via a singular gateway (Ramirez et. al, 2019). Figure 3 below outlines a typical star of stars LoRa topology.

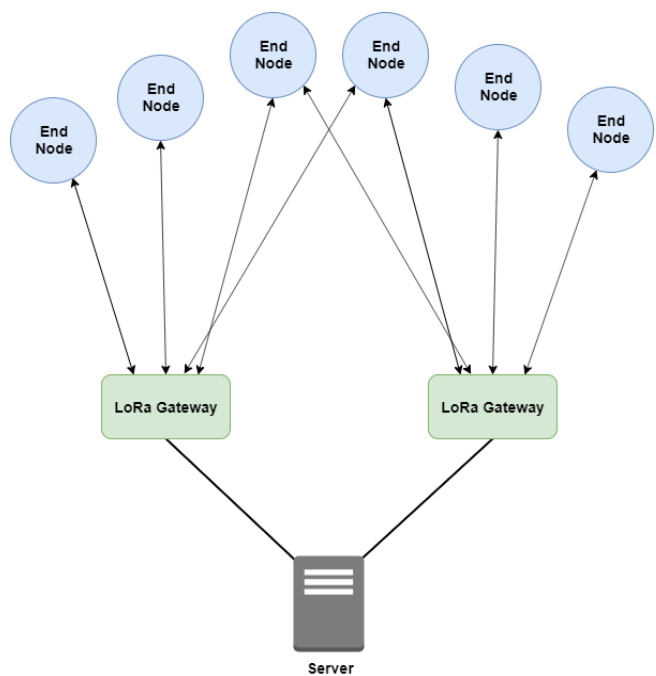

**Figure 3.** A LoRaWAN star-of-stars topology

This introduces issues as LoRa as a protocol is designed mostly to be offline most of the time to save power as it serves constrained devices, particularly those that operate in the Class A category. These devices conserve energy through this method, and end-devices cannot receive a message unless an uplink packet has been transmitted by the end-node. Class A devices are described as the lowest power end-device system, where applications can only receive data shortly after transmission (LoRa Alliance, 2017). Following this transmission, two receive windows are created: the first is one second long, as defined by the default value of *RxDelay* set to 1 second. The second window is opened after 1 second after the first window. In order to accurately receive transmissions on time, the uplink packet is required to be transmitted at a precise time to allow for the gateway and its limited throughput to serve each end-node. The current time request method of *DeviceTimeRequest* as defined by the LoRa specification has a time accuracy synchronisation requirement of +/-100ms. Compared to other time synchronisation methodologies, this is not remarkably accurate and the function itself does not account for drift of the internal clock.

1.5 Time Synchronisation Implementation Difficulties In LoRaWAN

Although the data transmission methods and capabilities have been outlined and researched extensively in various publications, the area of system TS within LoRa devices has only recently garnered attention. This is owed to the increasing importance of time-sensitive data transmissions within dispersed nodes in a network. Not only is TS in IoT in general lacking in research, but the current research surrounding TS accuracy and methods describe a number of factors that affect performance when attempting synchronisation. These include some general timing problems such as oscillator behaviour, timestamping accuracy, clock adjustment, and a lack of higher quality hardware components being utilised commercially (Mahmood et al, 2016, p. 4). Issues are further exacerbated through wireless nodes containing internal clocks that are subject to higher drift over time due to various manufacturing imperfections and environmental changes (Gonzalez, 2015). Swain & Hansdah (2015) describe the drift rate of a clock as follows, where $C_i(t)$ is the clock time of node *i at* real-time *t*. *p* is the drift rate of the clock at node *i*, and can be calculated with Equation 1.

$$\rho = \lim_{\Delta t \to 0} \frac{C_i(t + \Delta t) - C_i(t)}{\Delta t} - 1.$$

**Equation 1.** Drift rate calculation (Swain & Hansdah, 2015)

Typically, the drift is represented as units such as ppm (parts per million) or μs/s. It is also described that additionally if a clock's drift remains constant over a period of time, the clock time of $C_i(t)$ at node $i$ in real time $t$ can be calculated through Equation 2 below.

$$C_i(t) = (1 + \rho)t + b.$$

**Equation 2.** Clock time of clock at node $i$ at real-time $t$ calculation (Swain & Hansdah, 2015)

Additional difficulties encountered within sensor networks include the offset of a clock, synchronisation error, and the synchronisation interval. The offset of a clock compared to another clock will continue increasing unless the offset is reduced through modifying the clock values. Logical clock time is sometimes derived from the physical clock value rather than modifying the clock value directly. This leads to synchronisation error, which is defined as the difference in a logical clock time of any pair of clocks. Synchronisation intervals are also to blame for inaccurate time - if a synchronisation interval is not set with an appropriate interval, the chance of the internal clock drifting out of synchronicity is higher. LoRaWAN in its current state has a low-duty cycle with high latency, which translates to traditional TS protocols such as MQTT and SNTP not being suitable due to their bandwidth-heavy nature.

1.6 Aims & Objectives

Three solutions have been proposed that attempt to address the problem of precision of the device clocks. LongShoT: long-range synchronisation of time, Time Synchronisation for Wireless Sensors Using Low-Cost GPS Module and Arduino, and Time synchronization in vehicular ad-hoc networks: A survey on theory and practice. (Ramirez et. al, 2019; Koo et. al, 2019; Hasan et. al, 2018). The first paper LongShoT is in-scope for this area due to its direct relevance to LoRaWAN, whereas the other two are out of scope due to their target areas of wireless sensor networks (WSN) and vehicular ad-hoc networks (VANET). This systematic literature review aims to analyse each individual methodology and its findings in an attempt to bolster the time synchronisation methods currently available to LoRaWAN. It is clear that further research is necessitated to determine an ideal TS method to allow for IoT devices to efficiently synchronise time accurately. This research will centre around addressing the following two research questions:

- What are the obstacles encountered when implementing time synchronisation methods in LoRaWAN?

- What methodologies currently tested and available to other similar IoT protocols are able to be applied to LoRaWAN?

These result in the objectives of this literature review, where if accurate TS in such devices is achieved, this can lead to the following:

- Higher accuracy in data transmission and retrieval in IoT network infrastructure

- Improved coordination in dispersed nodes within an IoT network

This literature review aims to outline existing proposed methods for various IoT protocols and contrast them with each other in terms of benefits and weaknesses. This is achieved through structuring this review as follows – an outline of the project methodology, an overview of the necessity of synchronising time, a requirement analysis of TS in LoRaWAN, outlining available TS strategies in various wireless IoT verticals, and a characterisation of time synchronisation techniques in LoRaWAN, followed by the discussion and conclusion. It should be noted that although LoRaWAN is a very specific IoT protocol, the research within this review may be applied to other aspects of IoT as a whole.

## 2. Project Methodology

This section will focus on describing the project methodology that will be adopted throughout this project. The project approach will be justified, along with the tools and research methods that were used in order to extract the data. Four stages have been identified, and these stages have been presented as a waterfall to properly represent the waterfall project management approach.

As waterfall is a non-iterative process, each task must to be completed in full in chronological order before the next stage is to be attempted. These four phases have be described in detail, in addition to the evaluation of their strengths and weaknesses.

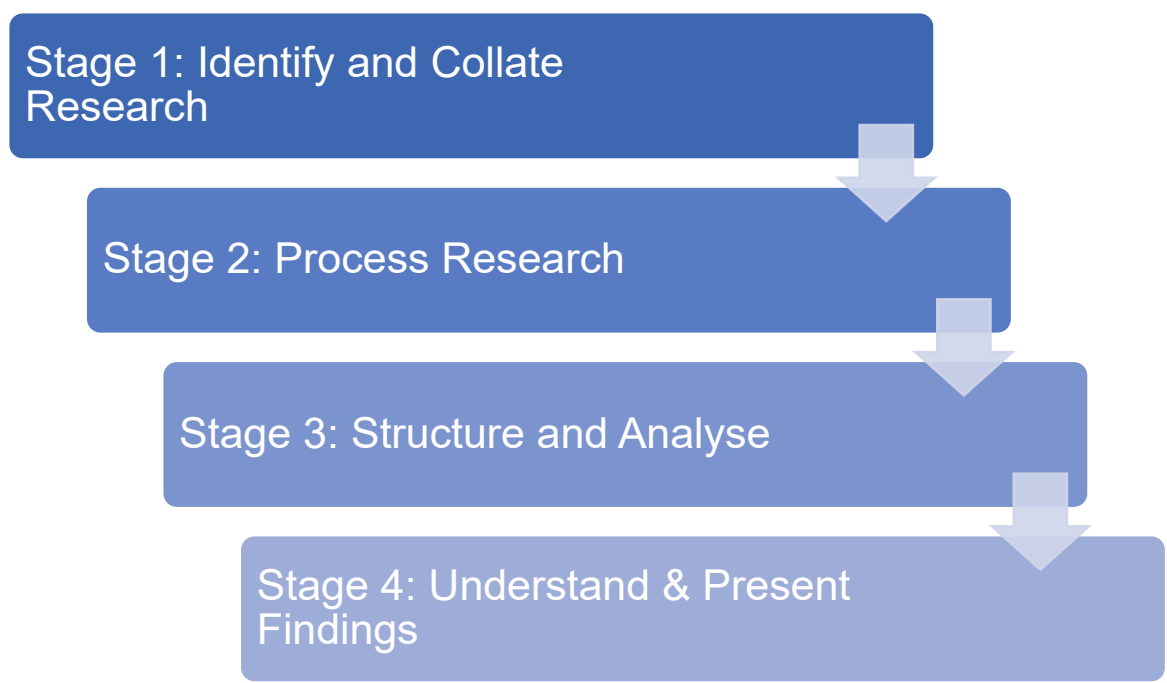

**Figure 4.** Four-stage methodology process

### Stage 1 – Identify and Collate

The project methodology approach is one of a data-oriented nature. Qualitative data will be extracted from a two-phase research approach, whereby a literature review will be performed on relevant articles and reports relating to time synchronisation in similar dispersed wirelessly networked devices. The literature was found through QUT's Library search portal, in combination with Google Scholar. Only research that was conducted and published within the years 2011 and 2019 was selected, so as to remove any non-useful studies that do not relate to LoRaWAN, given its recent inception. The literature that was obtained was not discriminated against by geographical origin, nor by original language. A keyword search approach was the adopted strategy in order to reveal relevant literature.

<u>Stage 2 – Process Research</u>

Immediately after the retrieval of relevant literature from the various searches performed within the previous phase, Stage 2 makes processes and categorises the literature. The articles, consisting of a total of 16, were added into the Nvivo software for further categorisation and analysis. Nvivo allowed for the identification of key themes and words to emerge and made identifying specifically relevant literature easy.

<u>Stage 3 – Structure and Analysis</u>

Following Stage 2, Stage 3 involves the creation of the Literature Matrix, where relevant literature was sorted into a table containing what identified them as relevant to the target topic of LoRa and Time Synchronisation. This allowed for quick viewing and refreshment of why the selected literature was added to the project's files and showed any out-of-scope information that may be contained within. Each study was arranged by its reference including the title, author, and DOI link, its relevancy to the project, and its correlation to the contexts of IoT and LoRaWAN.

<u>Stage 4 – Understand and Present Findings</u>

The final stage of the research methodology consists of the presentation of the findings, wherein the research data was extracted, comprehended, and formatted into an appropriately readable format. Through the structured approach of investigating each selected research area individually, it was found that the most efficient approach of literature analysis was to segment the literature. This meant analysing the strengths and weaknesses of each technology within each paper separately. As such, this literature review is structured in such a way that each TS platform belongs to its own section. The time period of this totalled five weeks, where Stages 1 to 3 consisting of the motivation and discovery of related TS methods in a relatively short time period of roughly a week. These past six weeks however have constrained the amount of literature that could possibly be analysed to under 100 articles, although it should be noted that despite this, LoRa TS is a quite new research field. This is reflected within the articles that cover other IoT technologies in place of LoRa, and although out-of-scope, aspects of each scheme can be applied to LoRa.

<u>3. The Necessity of Time Synchronisation</u>

Time synchronisation within IoT, specifically LoRaWAN, is highly important to allow for data consistency and coordination of devices within distributed systems. As most LoRaWAN nodes are tasked with obtaining sensor data, the collation of this data in an accurate manner is paramount. If one or more nodes are out of sync, the data being reported may be reported at an incorrect interval and be combined with the main datasets at the wrong interval. As further investment occurs into LoRaWAN as a low-power, long-range solution toward localisation and tracking, micro and even nano-second accuracy may be required in order to provide real-time data for analysis. A TS protocol that is better suited toward LoRaWAN is required, as although there exists traditional TS protocols, these are ill-suited toward LoRaWAN due to the protocol's strict bandwidth and power limitations.

Some use cases of LoRaWAN that outline precise TS requirements include smart metering and structural health monitoring (Laveyne et. al, 2017; Araujo et. al, 2011). Smart metering devices operate within Class B of the LoRa specification, whereby additional receive windows are opened at specified windows. It is described that highly accurate time synchronisation is required of the network for devices operating in the Class B mode, and this is confirmed in an experiment conducted in 2018 applying LoRa to smart metering. Four separate smart metering use cases of automated meter reading, time of use, outage monitoring, and quasi real-time monitoring were investigated, and their requirements specified. Their data transmission intervals ranged from 1 day to 1 minute, with the maximum throughput varying from 2.4 kbytes/day to only 1 byte per day. Of particular interest is the 1-minute latency figure, where if a LoRa device were to send data at irregular intervals due to a lack of proper TS, results will be skewed.

Structural health monitoring (SHM) on the other hand, is handled by wireless networks cover damage detection, damage localisation, damage quantification, and assessment of the remaining structural lifetime (Araujo et. al, 2012). Monitoring of structural health is important as it stems from the mitigation of potential hazards to the public caused by failing structures. Moving from traditional SHM setups which were restricted to wires caused new issues to arise, such as reliability, scalability, and time synchronisation accuracy especially. As with any sensor network with dispersed nodes, the clocks of each end-node will vary over time due to imperfections in oscillator crystals and ill-suited TS protocols. If the time is synchronised incorrectly or the node's internal clock is inaccurate, in the context of SHM applications this is highly dangerous due to any changes within the integrity of the structure being monitored not being reported at the precise time. TS within such a network in this case was achieved through the IEEE 802.15.4 IoT protocol for transmissions of time packets. This mitigates the need for a communication layer as an in-between.

4. Clocks in LPWAN

LoRa, part of the LPWAN class of networks, can have either internal or external clock sources when implementing the clock for TS. Examples of internal clock sources include crystal oscillators (CO), particularly temperature-compensated crystal oscillators (TXCO), with external clock sources being GNSS-based.

TXCO

Many vendors have opted to install a TXCO as the clock source for LoRa, due to the LoRa specification not outlining any specific type of oscillator to be used. This is most likely due to three improved characteristics in comparison to normal COs – high-precision, small-package size, and low power requirements. The PPM (parts per million) performance as defined by Rutkowski, R (2019) as being higher than a typical CO with up to 40 times better performance over a wide range of temperatures ranging from $0°c$ to $70°c$. The package itself is quite small, with average sizes of 5 x 3.2 x 1.5mm being highly suited to small LoRa end-nodes boards. Power requirements of TXCOs are also minute, as power draw can be as little as 2mA from power supplies of 3V, varying by manufacturer. It is also important that TXCOs have low-drift thermal characteristics in order to be an accurate clock source

GNSS-based

External clock sources based on GNSS (Global Navigation Satellite System, also commonly known as GPS) are usually placed within the gateway of a LoRa network (Barillaro et. al, 2019). This allows for hardware time stamping of received messages to occur in order to calculate the time to synchronise the end-nodes accurately. In some cases though, the GNSS module can also be placed directly onto the node for higher accuracy at the cost of power consumption. Due to the many satellite constellations available from various providers (Beidou, Galileo, GPS, GLONASS, QZSS), a higher accuracy can be achieved through GNSS time transfer techniques. These techniques involve using the GPS time from the satellites to estimate the real time, and these range from IGS timing, GNSS all-in-view (AV), GNSS common view (CV), two-way satellite time and frequency transfer (TWSTFT), GNSS carrier stage time transfer (CPTT), and laser synchronisation from stationary orbit (LASSO). Their accuracy can be viewed in the table below.

| GNSS Time Transfer Technique | Accuracy in ns | Availability | Application Area | General Technique |
|---|---|---|---|---|
| International GNSS Service (IGS) product timing | 0.03ns | 100% | Surveying, geomatics and geo-information | Un-differenced dual-frequency pseudo-range and carrier stage observations along with IGS precise orbit product |
| GNSS All in View (AV) | 1ns | 98% | International time comparisons | Absolute positioning for both stations using IGS precise ephemeris and clocks |
| GNSS Common View (CV) | 1-10ns | 98% | International time comparisons | Estimation of the integer ambiguities and clock difference |
| Two-way Satellite Time and Frequency Transfer (TWSTFT) | 1ns | 98% | NIST Automated Computer Time Service (ACTS) | Compare two signals that travel both ways between two clocks/oscillators |
| GNSS Carrier Stage Time Transfer (CPTT) | 0.3ns | 98% | International time comparisons | Adjusting bias using code measurements |
| Laser Synchronization from Stationary Orbit (LASSO) | 1ns | 98% | International time comparisons | Measures the intervals between arrival times of laser pulses synchronized with atomic time issued from multiple points on a single satellite |

**Figure 5.** Comparison of GNSS time transfer techniques

5. Requirements Analysis of TS in LoRaWAN

In the study conducted by Rizzi et. al (2017), LoRaWAN can be a solution to many distributed measurement applications (DMS). These DMS applications can range from industrial automation (factory & process), smart buildings (HVAC), home automation, smart metering, and smart grid. Although LoRaWAN is able to serve these various DMSs, each of these applications have different requirements ranging from update time, latency, sync error, data size and number of nodes within the LoRa network. This can be seen in the table located in Figure 2 below.

| Application | Update time (s) | Latency (s) | Sync err. (s) | Data Size (B) | # Nodes |
|---|---|---|---|---|---|
| Industrial Automation (Factory) | <0.1 | <0.01 | $<10^{-4}$ | <100 | $10^2$ |
| Industrial Automation (Process) | <60s | <1 | <0.01 | <100 | $10^3$ |
| Smart Building (e.g. HVAC) | <600 | <60 | <1 | <1k | $10^2$ |
| Home Automation (e.g. lighting) | Event based | <60 | $<10^{-4}$ | <100 | $10^2$ |
| Smart metering (Electricity) | <600 | <10 | <1 | 100 | $10^6$ |
| Smart metering (Gas) | 4/day | <3600 | <60 | 100 | $10^6$ |
| Smart Grid (PMU) | <0.01 | <0.01 | $<10^{-6}$ | <200 | $10^2$ |
| Smart Grid (EV fleet) | 4/day | <15 | <60 | <300 | $10^2$ |

**Figure 6.** The main requirements of typical DMSs. (Rizzi et. al, 2017)

Of particular interest is the update time, latency and sync error values in the table in seconds. Rizzi et. al concludes that despite the limited data rates specified in the LoRa protocols PHY and Data Link layers, it can meet the requirements for the DMSs. It is clear that although LoRa as a protocol can provide these DMS applications, various TS methodologies are required to suit each application. The data size in bytes and the number of nodes required for each application is also needed to be considered when applying a particular TS scheme to the LoRa network.

6. Overview of Available TS Strategies in the Various Wireless Networks

6.1 Hardware-Based – LongShoT

The LongShoT study aims to provide a solution to TS in LoRaWAN-based IoT devices. It attempts to do this through using deterministic properties contained within LoRaWAN networks, alongside hardware and MAC-level timestamping of packets. This research was conducted on the basis of LoRaWAN being designed with delays in transmission, low duty cycles, and wide deployment range, and not currently possessing a precise, bandwidth and power-sensitive method of TS. The proposed solution addresses this issue through sending a synchronisation transaction in the form of single packet transmission from the LoRa gateway.

The solution takes into account the propagation delay of up to 5μs between devices and the limitations of the LoRa PHY layer's chipping rates. The LongShoT protocol addresses these factors through utilising the gateway to send a small network request that can be embedded in packets such as telemetry. Following the response from the Network Server containing time information being received, the timestamp data is used to compute an accurate time. Prior to sending the packet, however, Two-Step Correction is applied to the LongShoT packet in order to account for propagation time and to estimate drift compared to the reference. This will be further described in the latter part of this section. It should be noted that although this single network request can transmit the packet to the node, the node receive window has to be open and ready after transmitting its uplink packet for the TS packet to be accepted. A typical transmission cycle for LoRaWAN Class A devices is shown in the figure below:

This is an unrefined manuscript that are under formatting for presenting as an Article Paper. This version of work is subjected to be replaced once the journal copy will be available.

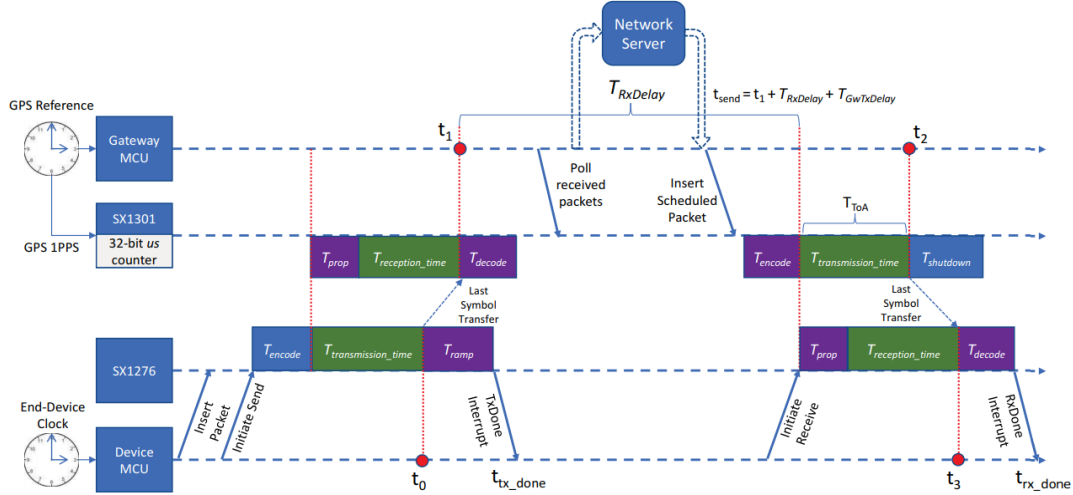

**Figure 7.** LoRaWAN Class A transmission cycle diagram (Ramirez et. al, 2019)

The results obtained from the short-range *OneShot* experiments consisting of devices located approximately 200m away from the gateway revealed that setting the *RxDelay* to 5 seconds allowed for sufficient time to correct drift rates to below 1ppm and bring synchronisation errors to within 3μs of GPS time. The long-range results of experiments where devices were located 1-4km away from the gateway resulted in drift rates two times as good as the natural drift rates usually exhibited by these devices. This is in addition to achieving synchronisation errors within ± 10μs. It is highlighted that this particular approach does not account for measurably comparing the accumulated drift with regards to the tick resolution of the device clock. Secondly, the difficulty of measuring accumulated drift against other delays of magnitude such as propagation delay may be high.

The second synchronisation method from LongShoT is the Long-Range, Two-Step Synchronisation method. Due to the long-range nature of the synchronisation request, there is a higher chance of error occurring with the request. This is where Two-Step Correction is introduced, which contains two network requests rather than one in order to estimate drift, whilst also accounting for propagation time and determining the exact clock offset. This addresses the shortcomings of the One-Shot short-range TS method, and begins by performing the first synchronisation request, which estimates the initial offset using time measurements. The second request is then made after a specified $T_{interval}$ to obtain secondary time measurements. The drift is then computed as according to the model below:

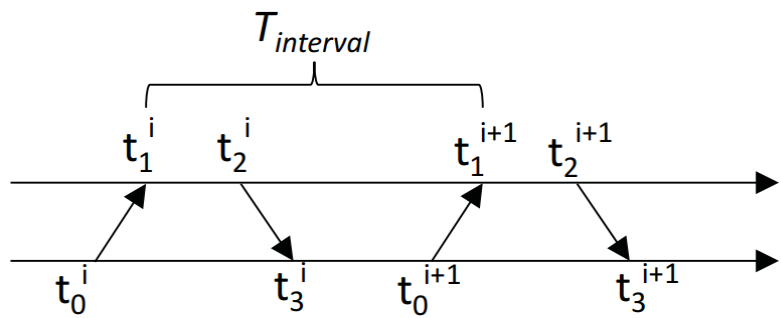

**Figure 8.** Two-Step Correction Model (Ramirez et. al, 2019)

In this method, the devices were programmed to synchronise with the gateway at 5-minute intervals for two hours. The end result was an offset from GPS time within ±100μs. This method is suited to perform adjustments taking into account long-term drift and aging of the oscillator. Advantages of this particular protocol in general also include the fact that drift rates and propagation delays are accounted for, where in traditional TS methods the time value from the Network Server is accepted without question. LongShoT also takes into consideration the limitations of the lower-powered hardware in off-the-shelf components (the 40MHz device crystal oscillators) and their propensity to create drift. The report outlines that although GPS accuracy in consumer-grade hardware is usually less than 100ns compared to 100μs, the power usage per sync is up to 2695mJ per sync (cold start). This is in comparison to the maximum transmission power of a 20dBm LongShoT request drawing only 190mJ of energy.

6.2 Hardware-based – Off-the-shelf GPS-Arduino in WSN

Koo et. al (2019) propose a two-stage solution for TS within Wireless Sensor Networks (WSN) to enable proper interpretation of measurements from such sensors. Measurement precision is important in wireless sensors as high sampling-frequency sensors provide a large amount of data, and this data is manipulated into mode shapes or strain mode shapes. Current studies surrounding RBS, FTSP and TPSN network-based synchronisation schemes used in WSNs are observed to provide time synchronisation with errors within 20$\mu s$. It is noted, however, that when root-based or tree-based topologies adopt these schemes, accumulative errors can reach up to 5*ms* in a period of 6 seconds. These errors have since been addressed through consensus-based time synchronisation, which doesn't require a single time reference.

Although the errors have been improved within these traditional TS methods, the more efficient method outlined is the use of a GPS module directly placed on each leaf node within the network in combination with a large-capacity battery pack. This approach has not been thoroughly investigated due to GPS power consumption widely acknowledged to be high in nature. The methodology indicated within the paper uses a low-cost GPS module alongside an Arduino Mega 2560 board with a combined cost of $80 USD. The synchronisation occurs through the resampling of two timestamps obtained through the input capture unit. The resampling output is determined by an equation where the inputs are the two timestamps and the measurement value of these.

The time-stamping method is comprised of two ideal PPS signals arriving at two separate intervals. These two signals can be also highlighted from the NMEA (National Marine Electronics Association) sentences that arrived before the two signals. The value of the first signal is read from the input capture unit and stored into a variable. The time value is identified as 1s later from the time located within the NMEA received sentence. The timer is then read when the *m*-th measurement is made, and steps one and two are repeated when the second signal arrives. Finally, the time of the *m*-th measurement by linear-interpolation is calculated, following Equation 3.

$$\hat{t}(m) = t_{P_I}(k) + \frac{C_{D_A}(m) - C_{P_A}(k)}{C_{P_A}(k+l) - C_{P_A}(k)} \times l \, (sec)$$

**Equation 3.** Estimated of measurement through linear interpolation (Koo et. al, 2019)

This resampling method of time-stamp comparison is shown in pseudocode as Algorithm 1.

```
Data: Timestamps t̂(m) and corresponding data y(m) for
      m = 0, 1, 2, ···
Result: Resampled data y_sync(q) on the regular
        timestamps t_sync(q) for q = 0, 1, 2, ···
initialization;
    t̂(0) ← the timestamp of the first measurement in
           second;
    y(0) ← the measurement value of the first
           measurement;
    t_sync(0) ← floor(t̂(0)) + 1; resampling starts from the
               exact start of the next second;
    m ← 1; measurement index;
    q ← 0; resampling index;
while True do
    t̂(m) ← the timestamp of the m-th measurement in
           second;
    y(m) ← the measurement value of the m-th
           measurement;
    if t̂(m − 1) ≤ t_sync(q) < t̂(m) then
        y_sync(q) =
            y(m − 1) + (t_sync(q) − t̂(m−1))/(t̂(m) − t̂(m−1)) × (y(m) − y(m − 1));
        t_sync(q + 1) ← t_sync(q) + ΔT_s; ΔT_s is the
                       resampling time interval;
        q ← q + 1;
    end
    m ← m + 1;
end
```

**Algorithm 1.** Pseudocode of implemented resampling accounting (Koo et. al, 2019)

The timestamping resampling method outlined within the paper is also subjected to error analysis. It is described that GPS receiver PPS signal time has an error offset. In the case of this particular GPS module, this offset time is 10*ns*.

Four sensors combined with the GPS modules were observed within the experiments. The number of satellites available to the GPS modules varied from 4 – 13. Although four experiments were performed, only results from Experiment #2 will be analysed as these are the specific results relating to time synchronisation accuracy. Results from this experiment which measured the accuracy of the PPS signals showed maximum deviations of 2 clock counts, which is equivalent to 200*ns*. This is highly accurate, and well within the requirements for high sampling-frequency sensors forming mode shapes and for data interpretation.

Accuracy of the time and availability of the GPS signals were high throughout testing, with recorded maximum relative time errors of 300*ns* for two adjacent PPS signals from a single module. Investigation of the proposed time stamping method revealed only two uncertainties – PPS signal errors and magnitude of the clock-period. This resulted in a maximum standard deviation of time stamping errors equalling 42.0*ns*.

6.3 Software-based – GNSS in VANET

The paper written by Hasan et. al (2018), describes the use of GNSS-based (Global Navigation Satellite System) technology within VANET-connected vehicles that currently use GNSS for positioning. Currently, it is said that GNSS, although mainly embraced for position data only, can be used in conjunction with existing time synchronisation protocols in order to accurately synchronise time. In today's VANETs, CSMA/CA (Carrier Sense Multiple Access/Collision Avoidance) is the selected medium to provide Time Synchronisation Function (TSF) data to vehicles. This is inadequate, as CSMA/CA is asynchronous only, and does not provide sub-second timing accuracy. Sub-second accuracy is necessitated within VANETs where network interoperability and coordination, scheduling of channels, road safety, security, and GPS-based relative vehicle positioning are required.

A solution consisting of receiving timing information from GNSS satellites through out-of-band external time synchronisation and using this to synchronise the VANET dispersed nodes located within the vehicles is proposed. The methodology is based on GNSS being highly stringent and accurate and can be relied on due to the information being generated from atomic clocks. As GPS receivers operating with GNSS are already inside most vehicles that are connected to a VANET, no additional hardware is required for this solution to work. The solution also outlays the fact that most new receivers are able to track multiple GNSS signals from various satellite GPS providers – namely GPS, GLONASS, Galileo and Beidou (BDS). This ensures connectivity at all times (up to 99.25% service with BDS + GPS combined) and also a comparison of time information for greater accuracy. A diagram showing how nodes individually synchronise via GNSS followed by updating with UTC time is displayed below:

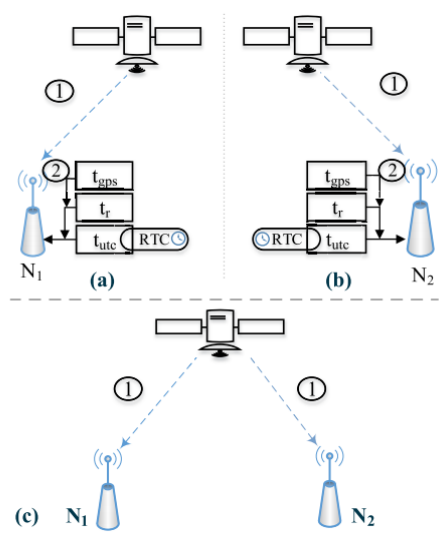

**Figure 9.** Nodes $N_1$ and $N_2$ individually synchronising with GNSS, updated with UTC (Hasan et. al, 2019, Hasan et. al, 2021)

Experiments conducted with this methodology employed two GPS receivers from U-Blox and Furuno with a clock containing an advanced temperature compensation circuit and quartz oscillator. Frequency stability of under 0.1ppm is specified for the oscillator. Accounting for different vendors adopting different error models and mitigation algorithms, a maximum time offset of 180ns was recorded between the two different models. This is in comparison to two of the same models being used, which yielded a time offset maximum of 60ns. Although these offsets exist, the timing errors were only within ± 10ns. This is acceptable and can be applied to use cases where timing accuracy is key. The graph of the time offset in ns between receivers of the same model can be viewed below:

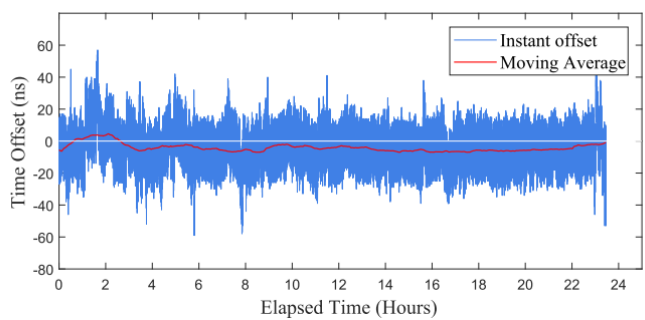

**Figure 10.** Results from same model GNSS vender over a long period (Hasan et. al, 2019)

This solution is beneficial in nature for VANET as it applies existing time synchronisation techniques such as TSF to existing hardware that are installed in connected vehicles. Given the fact that most GPS modules receive multiple signals from various GNSS providers, high accuracy and signal availability are also ensured. It is recommended however that in the interest of interoperability, the same vendor should be used in terms of GPS receiver. As most consumer-grade receivers are low-end in nature, this ensures that there are less time offsets between receiver time.

Following comparison of the three technologies, their accuracy, reliability and scalability have been determined. As they all utilise the native processing of the components on the node or gateway, albeit reprogrammed, their computational complexity is low. It should also be highlighted that the only methodology that added costs was the off-the-shelf GPS module added onto the leaf-nodes of a WSN. The LongShoT and GNSS-based methods only reprogram the hardware. The various statistics for each of the technologies analysed in the this section can be seen in the table below.

| Methodology | Technology | Accuracy | Reliability | Scalability | Computational complexity | Cost |
|---|---|---|---|---|---|---|
| LongShoT | LoRaWAN | 10μs | High | Yes | Low | $0 – reconfigures hardware |
| GNNS-based | VANET | 10ns | High | Yes | Low | $0 – reconfigures hardware |
| Off-the-shelf GPS | WSN | 200ns | High | Yes | Low | ~$80 USD (~$40 Arduino board + $40 GPS module) |

**Figure 11.** Comparison of the properties of LongShoT, GNSS, and Low-cost GPS TS methodologies

7. Characterisation of Time Synchronisation Techniques In LoRaWAN

7.1 Existing Work

7.11 A model for the classification and survey of clock synchronization protocols in WSNs

Previous research that has been conducted surrounding IoT technologies similar to that of LoRaWAN include the survey conducted by Swain et. al (2014), which outlines the need for accurate clock synchronisation within wireless sensor networks. It is said that although many clock synchronisation protocols have been introduced for use by WSNs, these protocols differ immensely from one another and suit various objectives. Protocols that are analysed such as Reference Broadcast Synchronisation (RBS), Timing Synchronisation Protocol for Sensor Networks (TPSN), Time Diffusion Synchronisation Protocol (TDP) are Flooding Time Synchronisation Protocol (FTSP) only represent a small amount of clock synchronisation protocols currently in use in WSNs. As these protocols each consist of a different mechanism of operation – for example computing offset versus averaging versus computing drift rate and offset – in conjunction with varying clock update and reading methods, it is unclear which of these schemes to select.

| Synchronization protocol | Fault tolerance | Scalability | Sleep scheduling | Energy efficiency | Global vs. local |
|---|---|---|---|---|---|
| RBS [20] | Yes | No | Yes | High | Local |
| TPSN [24] | Yes | Yes | Yes | Low | Global |
| TDP [16] | Yes | Yes | Yes | High | Global |
| FTSP [23] | Yes | Yes | Yes | Average | Global |
| Global clock synchronization [25] | Yes | Yes | No | Low | Global |
| GTSP [21] | Yes | Yes | Yes | Average | Global |
| Consensus based time synchronization protocol [19] | Yes | Yes | No | Low | Global |
| Probabilistic clock synchronization [28] | No | Yes | Yes | High | Global |
| Weighted average based clock synchronization protocol [22] | Yes | Yes | Yes | High | Global |
| External clock synchronization [17] | Yes | Yes | No | High | Global |

**Figure 12.** Global objective features of representative clock synchronisation protocols (Swain et. al, 2014)

A framework is presented by Swain et. al (2014) where ten of the main time synchronisation protocols are classified and their properties established. It can be seen from the framework that the energy efficiency is high for most of the synchronisation protocols outlined, and although dependent on the application and hardware requirements, most would be suited to LoRa.

7.12 LongShoT: Long-Range Synchronization of Time

Although there exists the previously mentioned frameworks containing synchronisation protocols currently offered to IoT, LongShoT is an individually developed protocol that does not rely on the built in *DeviceTimeRequest* function. This allows for a more fine-grain approach toward time synchronisation in LoRaWAN, in contrast to the study surrounding wireless sensor networks (WSN) only. The paper details the LoRa protocol and its limitations and strengths.

A need is outlined with regards to a LoRa-specific time synchronisation scheme that serves both short-range (~200m) and long-range synchronisation (>9km). Single-packet synchronisation over 200m is achieved through the One-Shot method which provides 10*ns* accuracy, as well as a slightly larger two-packet network request which achieves an offset of GPS time within ±100μs. This is one of the only LoRaWAN studies that have been conducted in recent times, and is highly suited for LoRa and various applications in its current state.

7.2 Proposed Method

A methodology has been created, based on the paper by Tirado-Andrés et. al (2019). The paper outlines a methodology proposed for selection of TS strategies in IoT in general; this altered model increases requirements and targets LoRaWAN specifically.

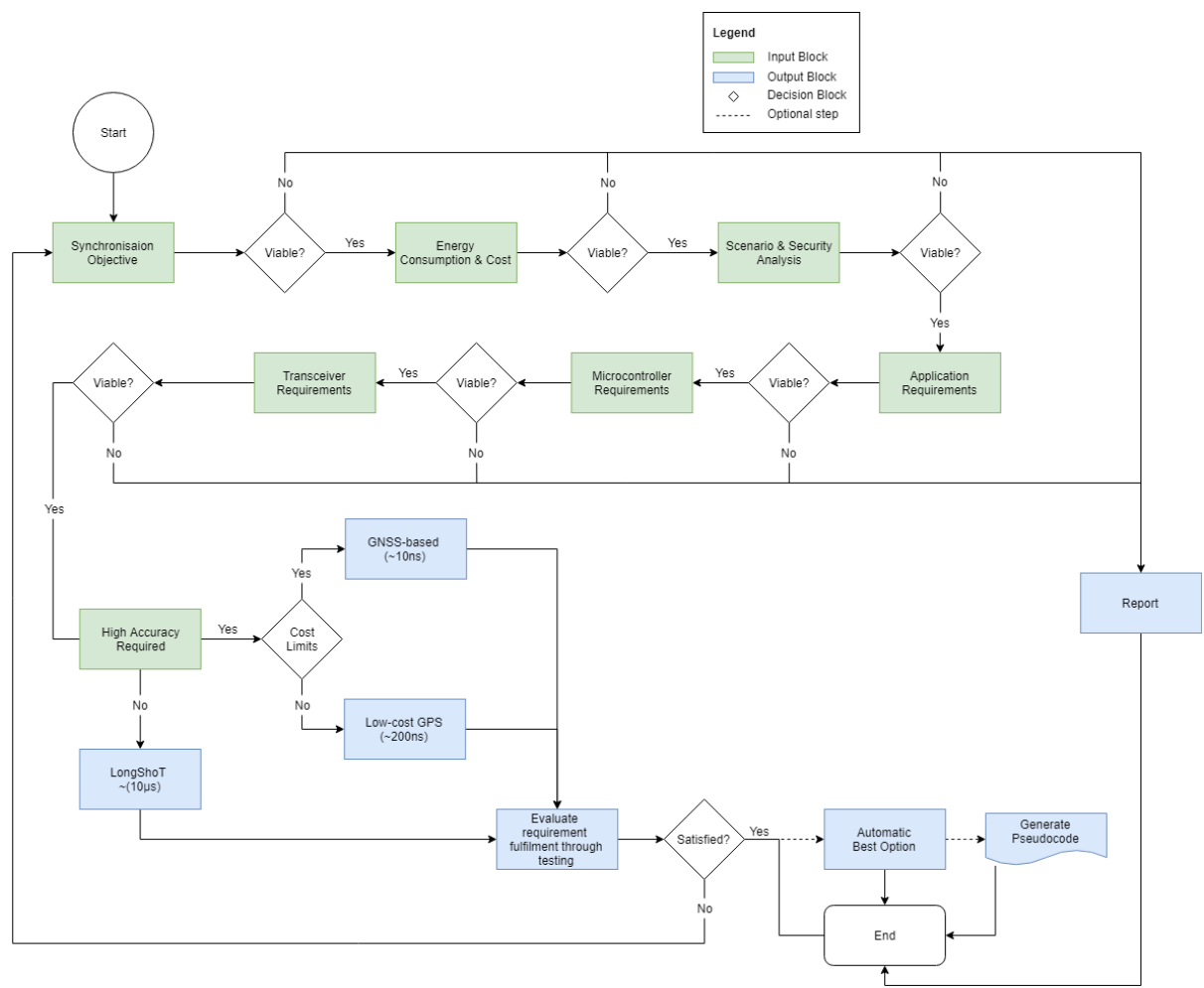

**Figure 13.** A refined model for selection of time synchronisation strategies in LoRaWAN (original by Tirado-Andrés et. al 2019)

This model borrows from the original through adopting typical flow behaviour, where each requirement influences the next requirement. Solutions are provided based on the following categories being viable: the synchronisation objective, energy consumption and cost, scenario and security analysis, application requirements, microcontroller requirements and transceiver requirements. If any of these are not viable, the next objective is to report, finishing with the end of the selection due to lack of a viable time-sync scheme being found. Each block is said to be interchangeable if developers are not satisfied with the outcome from the methodology due to one or two blocks not being outright viable.

The first block is the synchronisation objective, where the most common objectives are defined within such as MAC schemes, data fusion, or communications with cooperative transmissions. This is followed by the energy consumption and cost block, where the interest of the developer regarding additional cost is evaluated. The scenario is then evaluated pertaining to where the solution is to be deployed. This block also assists in determining whether the particular environment supports or hinders the tasks relating to time synchronisation, and how critical security is to the objective. Application requirements follow, where the specific accuracy required for the synchronisation strategy is input. This block is described as one of the most important requirement blocks due to the error limit that the network can tolerate being set. If this error limit is exceeded, the strategy is invalid. Finally, the microcontroller and hardware resources of the transceiver are collated to view whether the time synchronisation strategy is viable if a particular platform is to be used.

If each of the requirements is indeed viable, then the model pivots to whether very high accuracy is not required, and if not then the TS scheme of LongShoT is selected with 10μs. If it is required, then there is a cost limitation decision to be made. When cost is not limited, the higher cost GNSS-based time-transfer scheme is selected with a synchronisation accuracy of 10ns. If the cost is limited, the low-cost off-the-shelf GPS resampling scheme is selected.

8. Discussion

The outcomes of this project allow the detailed analysis of time synchronisation schemes that can be applied to the context of LoRaWAN. Relevant literature was sought out and found to be lacking in this research area. The benefits of existing time synchronisation protocols in their application to various IoT technologies were outlined in comparison to other protocols however, there seems to be a dearth of LoRa-specific TS schemes. The literature also did not outline any application of the currently offered time synchronisation protocols to LoRaWAN and their effect on nodes fulfilling their duty when using these.

This gap in knowledge is addressed by discovering that the LoRa component SX1301 contains a path to slave the chip to an external GPS 1PPS signal. This opens the LoRa node to multiple various implementations of GNSS -based time synchronisation methodologies, albeit at higher power consumption. The VANET, WSN and LoRa-specific solutions were all outlined within this project as a result and can be adapted to LoRa with small modifications.

Thus, the aim of the project is satisfied as the application GNSS-based solutions provide LoRa with TS methodologies accurate to nanoseconds. The other included literature did not satisfy the requirements of the scope, although some information was obtained from each of these in order to better provide background on the subject matter. Although there were three individual strategies identified within the project, it is clear that further research is necessary in order to ascertain a native TS scheme relating to LoRa. The discovery of the relevant literature and its analysis applied throughout the report supplies motivation for further analysis and creation of time synchronisation strategies for LoRaWAN.

In comparison to the strengths of this report, some weaknesses are also identified. The first is that the strategies selected for time-synchronisation may not fulfilling the application requirements of most LoRa use cases where power is highly limited. High power consumption is built into in the design of GNSS, even when used sparingly. This limitation is owing to the fact that there does not currently exist an adequate cover-all solution that can be used for all LoRaWAN setups. A greater weakness is that this report only outlines the potential benefits of adopting the three specified strategies yet provides no physical testing with regards to the application to LoRa nor confirm any findings. It is recommended that testing be completed of each strategy applied to diverse LoRaWAN contexts. This will further divulge any inconsistencies when applying GNSS to LoRa, and also evaluate LongShoT in a long-term use-case. Despite this weakness, the need for additional research has been identified which allows practitioners to pursue and produce efficient time synchronisation schemes that may be able to be scaled to all IoT devices.

9. Conclusion

This paper, in the context of LoRaWAN, aimed to determine an appropriate time synchronisation strategy based on the application requirements of a LoRa network. It is significant due to the lack of current research surrounding LoRaWAN and synchronising time in general, aiding in increasing the efficiency of the protocol. This report identifies key elements such as the need for TS in LoRa, the limitations currently faced and how they may be overcome through three individually inspected technologies. These included using a LoRa-specific strategy in the form of LongShoT, a WSN-based solution using off-the-shelf GPS-modules and programmed Arduino boards, and a GNSS-based solution that uses GNSS time transfer in vehicular ad-hoc networks (VANET).

Furthermore, a methodology was adapted from a recent paper with LoRa as the selected IoT medium. A prospective developer utilising LoRaWAN in a new network may adapt the methodology by providing their own inputs into the various requirement blocks. Following each requirement being viable, the developer is presented with a selection of time synchronisation methods based on their needs.

This methodology in conjunction with the previously mentioned TS methods provide a baseline for an understanding of LoRaWAN's technological structure, its limitations, and limitations in IoT in general in addition to possible solutions. Although not all solutions aligned with the target IoT technology, findings extracted from each of the studies may provide developers and practitioners with practical information regarding how best to approach time synchronisation in LoRaWAN and IoT as a whole.